\documentclass[prd,twocolumn,floatfix,preprintnumbers]{revtex4}

\usepackage{graphicx}
\usepackage{amsmath}
\usepackage{graphicx,epsfig}
\usepackage{bm}
\usepackage{latexsym,amssymb,amsmath,float}

\newcommand {\ga} {\ {\raise-.5ex\hbox{$\buildrel>\over\sim$}}\ }
\newcommand {\la} {\ {\raise-.5ex\hbox{$\buildrel<\over\sim$}}\ }

\def\be{\begin{equation}}
\def\ee{\end{equation}}
\def\ba{\begin{eqnarray}}
\def\ea{\end{eqnarray}}

\begin{document}

\title{Dark energy with $w \rightarrow -1$: Asymptotic $\Lambda$ versus pseudo-$\Lambda$}
\author{Robert J. Scherrer}
\affiliation{Department of Physics and Astronomy, Vanderbilt University,
Nashville, TN  ~~37235}

\begin{abstract}
If the dark energy density asymptotically approaches a nonzero constant, $\rho_{DE} \rightarrow \rho_0$, then its equation
of state parameter $w$ necessarily approaches $-1$.
The converse is not true; dark energy with
$w \rightarrow -1$ can correspond to either $\rho_{DE} \rightarrow \rho_0$
or $\rho_{DE} \rightarrow 0$.  This provides a natural division of
models with $w \rightarrow -1$ into two distinct classes:
asymptotic $\Lambda$ ($\rho_{DE} \rightarrow \rho_0$) and pseudo-$\Lambda$ ($\rho_{DE} \rightarrow 0$).
We delineate the boundary between these two classes of models in terms of the behavior
of $w(a)$, $\rho_{DE}(a)$, and $a(t)$.  We examine barotropic and quintessence realizations of both types
of models.
Barotropic models with positive squared sound speed and $w \rightarrow -1$ are always asymptotically $\Lambda$;
they can never produce pseudo-$\Lambda$ behavior.  Quintessence models can correspond to either asymptotic
$\Lambda$
or pseudo-$\Lambda$ evolution, but the latter is impossible when the expansion is dominated by a background
barotropic fluid.
We show that the distinction between asymptotic $\Lambda$ and pseudo-$\Lambda$ models for $w> -1$ is mathematically dual to the distinction
between pseudo-rip and big/little rip models when $w < -1$.

\end{abstract}

\maketitle

\section{Introduction}

Cosmological observations \cite{union08,hicken,Amanullah,Union2,Hinshaw,Ade,Betoule}
indicate that roughly
70\% of the energy density in the
universe is in the form of a negative-pressure component,
called dark energy, with roughly 30\% in the form of nonrelativistic matter.
The dark energy component can be parametrized by its equation of state parameter, $w$,
defined as the ratio of the dark energy pressure to its density:
\be
\label{w}
w=p_{DE}/\rho_{DE}.
\ee
A cosmological constant, $\Lambda$, corresponds to the case, $w = -1$ and $\rho = constant$.

While a model with a cosmological constant and cold dark matter ($\Lambda$CDM) is consistent
with current observations,
there are many other models of dark energy that have a dynamical equation
of state.
For example, one can consider quintessence models, with a time-dependent scalar field, $\phi$,
having potential $V(\phi)$
\cite{RatraPeebles,Wetterich,Ferreira,CLW,CaldwellDaveSteinhardt,Liddle,SteinhardtWangZlatev}.
(See Ref. \cite{Copeland1} for a review), or barotropic models, in which the pressure is a specified
function of the density
\cite{Kamenshchik,Bilic,Bento,linear1,linear2,AB,Quercellini,VDW1,VDW2,LinderScherrer,Bielefeld},
as well as numerous other possibilities.
However, any of these models must closely mimic $\Lambda$CDM in order to be consistent
with current observations; in particular, any viable model should have a present-day value
of $w$ close to $-1$.

Given these current observational constraints on $w$, it is an opportune time
to examine more carefully the general properties of models in which $w \rightarrow -1$ asymptotically.
The most
straightforward
case is simply the group of models that asymptote to a nonzero constant value of $\rho_{DE}$, i.e,
that are asymptotically identical to $\Lambda$CDM, a group of models we will
call ``asymptotic $\Lambda$."  However, this is not the only possibility; there is a second class of models
for which $w \rightarrow -1$ and $\rho_{DE} \rightarrow 0$.  We will dub these ``pseudo-$\Lambda$"
models
because they represent the closest one can approach $\Lambda$ without the dark energy density
itself being asymptotically constant.  Note that this distinction is independent of the underlying physical
model for dark energy; it is simply a statement about the asymptotic evolution of $\rho_{DE}$ and $w$.

Pseudo-$\Lambda$ models are not new; for example, they can be the natural end state
of ``freezing" quintessence models, and some classes of them correspond to
previously-investigated models for inflation.  However, there has been to date no systematic study of the
boundary between pseudo-$\Lambda$ and asymptotic $\Lambda$ models, or a discussion of the way this distinction
is realized in physical models.  These are the main aims of this paper.

In the next section, we explore, in turn, the conditions on $w(a)$, $\rho_{DE}(a)$, and $a(t)$, where
$a$ is the cosmological scale factor and $t$ is time, that determine whether
a given model will yield asymptotic $\Lambda$ or pseudo-$\Lambda$ evolution.  In Sec. III we determine the conditions
on two classes of physical models (barotropic models and quintessence models) corresponding to either type of behavior.
In Sec. IV, we examine the relation between the asymptotic $\Lambda$ and pseudo-$\Lambda$
models for $w > -1$ and the pseudo-rip and big/little rip models for $w < -1$.  We discuss our results in Sec. V.
We will take $\hbar = c = 1$ throughout and work in units
for which $8 \pi G = 1$.

\section{Distinguishing asymptotic $\Lambda$ and pseudo-$\Lambda$ models}

Consider dark energy that evolves asymptotically in
one of the following three ways:
\begin{align*}
&{\rm (I)} ~~~~w \rightarrow -1, ~~~~~~~~~\rho_{DE} \rightarrow \rho_0 \ne 0, ~~ ({\rm asymptotic}~\Lambda)\\
&{\rm (II)} ~~~w \rightarrow -1, ~~~~~~~~~\rho_{DE} \rightarrow 0,  ~~~~~~~~~({\rm pseudo-}\Lambda)\\
&{\rm (III)} ~~w \rightarrow w_0 \ne -1, ~~\rho_{DE} \rightarrow 0. 
\end{align*}
Type I corresponds to asymptotic $\Lambda$ evolution, while type II represents pseudo-$\Lambda$ models.
Type III corresponds to models which do not have $w \rightarrow -1$, but such models can be made consistent
with observations if $w_0$ is sufficiently close to $-1$.  Our goal is to delineate the boundaries
between these three types of behavior, first in terms of conditions on $w(a)$, second
in terms of $\rho_{DE}(a)$, and finally in terms of the behavior of $a(t)$.

\subsection{Specified $w(a)$}

Consider first the case where we specify $w$ as a function of $a$.
The evolution of the dark energy density as a function of the scale factor $a$ is given by
\begin{equation}
a \frac{d\rho_{DE}}{da} = -3(\rho_{DE} + p_{DE}).
\end{equation}
It is convenient to rewrite this in terms of the quantity $w$:
\begin{equation}
\label{evolve}
\frac{d \ln \rho_{DE}}{d \ln a} = -3(1+w).
\end{equation}
It is now straightforward to derive the conditions on $w$ as a function of $a$ that
correspond to the three types of evolution defined above.

Consider first the boundary between type II and type III.  This is simply determined
by the condition $w(a) \rightarrow -1$; when this condition is satisfied, the dark energy
evolves as in type I or II, while $w \rightarrow w_0 \ne -1$ corresponds to type III
evolution. The boundary between asymptotic $\Lambda$ (type I) and pseudo-$\Lambda$ (type II) can be determined by integrating
Eq. (\ref{evolve}):
\begin{equation}
\label{integral}
\ln \rho_{DE} = -3 \int (1+w)~ d \ln a.
\end{equation}
Note that we are interested in the asymptotic (large-$a$) behavior of $\rho_{DE}$, so we can ignore
the behavior of the integral at small $a$.
It is clear that $\rho_{DE} \rightarrow \rho_0$ (type I, asymptotic $\Lambda$) when the integral in Eq. (\ref{integral})
converges as $a \rightarrow \infty$, while $\rho_{DE} \rightarrow 0$ corresponds to divergence
of the integral (type II, pseudo-$\Lambda$).

We now have the conditions on $w(a)$ to produce pseudo-$\Lambda$ behavior; this requires
\begin{equation}
w \rightarrow -1,
\end{equation}
as $a \rightarrow \infty$
and
\begin{equation}
\label{intcondition}
\int_{a_0}^\infty (1+w)~ d\ln a \rightarrow \infty.
\end{equation}

To illustrate these results, let us consider an equation of state parameter given, in the limit of large $a$, by
\begin{equation}
\label{gammaexample}
1+w = \frac{A}{(\ln a)^q},
\end{equation}
where $A$ and $q$ are constants.
This satisfies the conditions for pseudo-$\Lambda$ evolution as long as $0 < q \le 1$.  When $q = 0$, we have, instead,
nonzero constant $w$ (type III), while $q > 1$ evolves to asymptotic $\Lambda$.
We can integrate Eq. (\ref{evolve}) to derive $\rho(a)$
for the pseudo-$\Lambda$ cases corresponding to Eq. (\ref{gammaexample}); we obtain (in the asymptotic limit of large $a$)
\begin{equation}
\label{rhoq}
\rho_{DE} \sim e^{ -\frac{3A}{1-q}(\ln a)^{1-q}} \sim a^{-\frac{3A}{1-q}(\ln a)^{-q}}
\end{equation}
for $0 < q < 1$, while $q=1$ gives
\begin{equation}
\label{logmodel}
\rho_{DE} \sim (\ln a)^{-3A}.
\end{equation}
Of the pseudo-$\Lambda$ models corresponding to Eq. (\ref{gammaexample}), the model closest to $\Lambda$CDM,
in the sense of having the most slowly decaying density, is the model with the most rapidly-decaying $w$, i.e.
the $q=1$ model.  Conversely, the most rapidly evolving $\rho_{DE}$ corresponds to the limit $q \rightarrow 0$.

There are, however, no sharp boundaries between the pseudo-$\Lambda$ models and those behaving as types I
and III, in the sense that for any given pseudo-$\Lambda$ model, one can always find another pseudo-$\Lambda$ model
for which
$\rho_{DE}$ decays more slowly (closer to type I) or more rapidly (closer to type III).  So, for instance, instead of the model described by Eq. (\ref{gammaexample})
with $q=1$, we can take
\begin{equation}
\label{border1}
1+w = \frac{A}{(\ln a)(\ln_2 a)(\ln_3 a)...(\ln_m a)}
\end{equation}
where we have defined $\ln_j(x) \equiv \ln \ln \ln ...\ln(x)$, with the logarithm on the
right-hand side iterated $j$ times.
This yields a value for $\rho_{DE}$ that declines extraordinarily slowly with $a$:
\begin{equation}
\label{logrho}
\rho_{DE} \sim (\ln_m a)^{-3A},
\end{equation}
If, however, we take instead
\begin{equation}
\label{border2}
1+w = \frac{A}{(\ln a)(\ln_2 a)(\ln_3 a)...(\ln_m a)^{1+\epsilon}},
\end{equation}
where $\epsilon > 0$ is a constant,
then  the integral in Eq. (\ref{integral}) converges regardless of how small $\epsilon$ is (cf. Ref. \cite{lrip}),
and $\rho_{DE}$ asymptotes to a nonzero constant (asymptotic $\Lambda$).
Given the very slow rate of growth of the function $\ln
_m a$ for large $m$, Eqs. (\ref{border1}) and (\ref{border2}) provide
a practical boundary between pseudo-$\Lambda$ and asymptotic $\Lambda$ behavior, although of course one can always
derive a form for $w(a)$ lying between these two functions that displays either kind of behavior.

At the other boundary, between pseudo-$\Lambda$ and type III, we have already noted that $q > 0$
in Eq. (\ref{gammaexample}) can be arbitrarily small
for pseudo-$\Lambda$ models; for any given value of $q$, one can always take $q$ to be smaller and obtain
a model for which $\rho$ decays more rapidly as a function of $a$.

\subsection{Specified $\rho_{DE}(a)$}

Now suppose instead that we specify the density, $\rho_{DE}$, as a function of $a$.  In this case, the condition for asymptotic $\Lambda$ (type I) evolution is trivial;
by definition it corresponds to $\rho_{DE}(a) \rightarrow \rho_0 \ne 0$ as $a \rightarrow \infty$.  On the other hand
$\rho_{DE}(a) \rightarrow 0$ can correspond to either pseudo-$\Lambda$ or type III behavior, so we need to distinguish
the conditions for these two types of evolution.

The density as a function of $a$ can always be written in terms of a function $f(x)$ in the somewhat unusual form
\begin{equation}
\label{rhoa}
\rho_{DE} \sim e^{-f(\ln a)}.
\end{equation}
Now consider the conditions necessary for pseudo-$\Lambda$ behavior.  In order for $\rho_{DE} \rightarrow 0$ asymptotically,
we must have $f(x) \rightarrow \infty$ as $x \rightarrow \infty$.  However, we also need $w \rightarrow -1$.  From our definition
in Eq. (\ref{rhoa}) and Eq. (\ref{evolve}), we will have $w \rightarrow -1$ as long as $f^\prime(x) \rightarrow 0$ for
$x \rightarrow \infty$.

This gives us the conditions on $\rho_{DE}(a)$ for pseudo-$\Lambda$ behavior, namely, any $f(x)$ satisfying
\begin{align}
\label{f1}
&f(x) \rightarrow \infty,\\
\label{f2}
&f^\prime(x) \rightarrow 0,
\end{align}
as $x \rightarrow \infty$ will generate a pseudo-$\Lambda$ model with $\rho_{DE}$ given by Eq. (\ref{rhoa}).

The two simplest functions satisfying Eqs. (\ref{f1})-(\ref{f2}) are $f(x) = x^{\alpha}$ with $0 < \alpha < 1$
and $f(x) = \alpha \ln(x)$, with $\alpha>0$, which correspond precisely to the functional forms for $\rho_{DE}(a)$ in Eqs.
(\ref{rhoq}) and (\ref{logmodel}), respectively.  However, these are, of course, just two of the many forms for $\rho_{DE}(a)$
that can be derived from Eqs. (\ref{rhoa}) - (\ref{f2}).

\subsection{Specified $a(t)$}

Asymptotic $\Lambda$ and pseudo-$\Lambda$ models can also be expressed in terms of the behavior of the scale factor $a$ as a function of the time $t$.
For a spatially-flat universe, the Friedman equations are
\begin{align}
&\left(\frac{\dot a}{a}\right)^2 = \frac{\rho}{3}, \\
&\frac{\ddot a}{a} = - \frac{1}{6}(\rho + 3 p).
\end{align}
Now take the expansion factor to be given in terms of $t$ as
\begin{equation}
a = e^{f(t)}.
\end{equation}
Substituting this expression into the Friedman equations and using the definition of $w$, we obtain
\begin{equation}
\rho_{DE} = 3 \dot f^2
\end{equation}
and
\begin{equation}
\label{gammaf}
1+w = - \frac{2 \ddot f}{3 \dot f^2},
\end{equation}
where we have assumed here that the expansion is dominated asymptotically by the dark energy.
We can now express our conditions for types I-III evolution in terms of $f(t)$ and its derivatives:
\begin{align}
\label{fcase1}
&{\rm (I)}  ~~~~\dot f \rightarrow constant \ne 0, ~~~\ddot f/\dot f^2 \rightarrow 0, \\
\label{fcase2}
&{\rm (II)} ~~~~\dot f \rightarrow 0, ~~~\ddot f/\dot f^2 \rightarrow 0,  \\
\label{fcase3}
&{\rm (III)} ~~\dot f \rightarrow 0, ~~~\ddot f/\dot f^2 \rightarrow constant \ne 0,
\end{align}

For asymptotic $\Lambda$
models (type I), one always has asymptotic de Sitter evolution,
\begin{equation}
a \sim e^{\sqrt{\rho_0/3}~t}.
\end{equation}
Similarly, models which asymptote to a constant non-$\Lambda$ equation of state (type III) give the standard
result
\begin{equation}
a \sim t^{2/3(1+w_0)}.
\end{equation}
In contrast, pseudo-$\Lambda$ models (type II) yield a wider variety of asymptotic behaviors for $a(t)$, some
of which have been explored previously in the context of inflation.

In general, any $f(t)$ satisfying Eq. (\ref{fcase2}) will correspond to a pseudo-$\Lambda$ model.  Consider
first the restricted class of such functions examined by Barrow \cite{Barrow}, who pointed out that
whenever $\ddot a$ is a rational function of $a$ and $t$, there are only a limited number of asymptotic behaviors
possible for $a(t)$.  For the models examined here, this corresponds to the case where $\rho_{DE}(a)$ and $w(a)$ are rational
functions of $a$.  Of the asymptotic behaviors examined in Ref. \cite{Barrow}, only two correspond to pseudo-$\Lambda$
behavior, namely
\begin{equation}
\label{intermediate}
a \sim \exp(A t^\alpha),
\end{equation}
with $0 < \alpha < 1$, and
\begin{equation}
\label{logamediate}
a \sim \exp[A (\ln t)^{\alpha}],
\end{equation}
with $\alpha > 1$, where $A$ is a constant in both cases.
These represent, respectively, intermediate inflation \cite{inter1,inter2,inter3} and logamediate inflation \cite{logbarrow}.

Using Eq. (\ref{gammaf}), one can work backwards to derive the corresponding $w(a)$.  For intermediate
inflation (Eq. \ref{intermediate}) we obtain
\begin{equation}
1+w = \frac{2}{3}\left(\frac{1}{\alpha} - 1\right)\frac{1}{\ln a}.
\end{equation}
This is simply the model presented earlier in Eq. (\ref{gammaexample})
with $q=1$.  For logamediate inflation (Eq. \ref{logamediate}) we obtain
\begin{equation}
1+w = \frac{2}{3 \alpha} A^{-1/\alpha} (\ln a)^{1/\alpha -1} - \frac{2}{3}\left(1 - \frac{1}{\alpha}\right)\frac{1}{\ln a}.
\end{equation}
Although the first term is of the same form as in Eq. (\ref{gammaexample}), the addition of the second term yields a model slightly different
from the one examined earlier.  While these two models exhaust the possibilities for asymptotic pseudo-$\Lambda$ behavior
when $w$ and $\rho_{DE}$ are rational functions of $a$, the latter condition is quite restrictive and will not apply to most
cases.

It is straightforward to derive the behavior of $a(t)$ for pseudo-$\Lambda$ models given in terms
of $\rho_{DE}(a)$; we have simply
\begin{equation}
t = \int \sqrt{\frac{3}{\rho_{DE}(a)}} d \ln a.
\end{equation}
For example,
for the density evolution in Eq. (\ref{logrho}), one can find
an exact solution for $m = 2$ and $A = 2/3$, namely
\begin{equation}
\label{t(a)}
t = \sqrt{3}[\ln a (\ln \ln a - 1)].
\end{equation}
While diverging (as expected) from de Sitter
expansion, Eq. (\ref{t(a)}) is manifestly ``closer" to the de Sitter expansion
law
($t \sim \ln a$)
than is the intermediate inflation expansion law ($t \sim (\ln a)^{1/\alpha})$.

\section{Physical Realizations}

In this section we will examine some representative physical models for dark energy to determine when such models
exhibit asymptotic $\Lambda$ behavior or pseudo-$\Lambda$ behavior.  In particular, we will investigate barotropic dark energy
models,
in which the pressure is a specified function of the density, and quintessence models, in which the dark energy arises
from a minimally-coupled scalar field.  This is by no means an exhaustive list of possibilities; one could also examine
$k$-essence models, noniminally coupled scalar fields, and a variety of other models.  However, the two classes
of models discussed here are among the simplest and most widely-studied, and our results will provide some insight
into the conditions needed for each type of $w \rightarrow -1$ behavior.  Our methodology can easily be extended to
other classes of dark energy models.

\subsection{Barotropic models}

Here we examine barotropic models, for which the pressure is a fixed function of the density:
\begin{equation}
\label{baro}
p_{DE} = f(\rho_{DE}).
\end{equation}
Particular models of this form include
the Chaplygin gas \cite{Kamenshchik,Bilic} and 
the generalized Chaplygin gas \cite{Bento}, the linear equation of state 
\cite{linear1,linear2} and the affine equation of state \cite{AB,Quercellini}  
(note these are actually the same model), the quadratic 
equation of state \cite{AB}, and the Van der Waals equation of state
\cite{VDW1,VDW2}.  A general study of the properties of barotropic models for dark energy
was undertaken in Ref. \cite{LinderScherrer} and further extended in Ref. \cite{Bielefeld}.

It is convenient to rewrite Eq. (\ref{baro}) in the form
\begin{equation}
\label{betterbaro}
p_{DE} = - \rho_{DE} + g(\rho_{DE}),
\end{equation}
where the function $g$ completely specifies the barotropic model.
In terms of our previous discussion, we have $1+w = g(\rho_{DE})/\rho_{DE}$, where $1+w$ is now
given as a function of the density.  For asymptotic $\Lambda$ behavior, the
density evolves to the constant value of $\rho_0$ with $w \rightarrow -1$, so we must
have $g(\rho_0) = 0$ for some nonzero $\rho_0$. Then $g(\rho_{DE})/\rho_{DE} \rightarrow 0$ as
$\rho_{DE} \rightarrow \rho_0$.  For pseudo-$\Lambda$ models, in contrast, we have
$g(\rho_{DE})/\rho_{DE} \rightarrow 0$ as
$\rho_{DE} \rightarrow 0$.

Now consider an example of each type of model.  In the generalized
Chaplygin gas model \cite{Bento}, the pressure as a function of density is given by
\begin{equation}
\label{baropower}
p_{DE} = - \frac{A}{\rho_{DE}^\alpha}.
\end{equation}
Then $g(\rho_{DE}) = \rho_{DE} - A/\rho_{DE}^\alpha$
and $g(\rho_{DE})/\rho_{DE} = 1 - A/\rho_{DE}^{\alpha +1}$.
We see that both $g(\rho_{DE})$ and $g(\rho_{DE})/\rho_{DE}$
go to zero
when $\rho_{DE} = A^{1/(\alpha + 1)}$.
Thus, the Chaplygin gas behaves as an asymptotic $\Lambda$ model, evolving to the constant nonzero density
$\rho_{DE} = A^{1/(\alpha + 1)}$.

In contrast, consider the equation of state
\begin{equation}
\label{quadratic}
p_{DE} = -\rho_{DE} + A \rho_{DE}^{\alpha},
\end{equation}
which was examined in Refs. \cite{Od1,Od2,Stef,Noj} (and in Ref. \cite{AB} for the special case $\alpha = 2$).
We have
$g(\rho_{DE})/\rho_{DE} = A \rho_{DE}^{\alpha - 1}$, and we see
that $g(\rho_{DE})/\rho_{DE} \rightarrow 0$ as $\rho_{DE} \rightarrow 0$ as long as $\alpha > 1$.
Thus, this model with $\alpha > 1$
corresponds to a pseudo-$\Lambda$ model.

However, there is an additional
condition that we can impose
on barotropic models.  Linder and Scherrer \cite{LinderScherrer} emphasized that 
stability of the dark energy fluid  requires that the sound speed, which is given by
\begin{equation}
c_s^2 = \frac{dp_{DE}}{d\rho_{DE}},
\end{equation}
should satisfy $c_s^2 \ge 0$.
Then the equation of state function in Eq. (\ref{baro})
is constrained to satisfy $df/d\rho_{DE} \ge 0$, which translates into a constraint on the function $g(\rho_{DE})$ in Eq.
(\ref{betterbaro}) of $dg/d\rho_{DE} \ge 1$.

Now consider pseudo-$\Lambda$ behavior.
This requires
$g(\rho_{DE})/\rho_{DE} \rightarrow 0$ as $\rho_{DE} \rightarrow 0$. Note that if $g(\rho_{DE})/\rho_{DE} \rightarrow 0$, then $dg/d\rho_{DE} \rightarrow 0$
(by L'Hopital's rule), so $c_s^2 = -1$.  Thus, for $c_s^2 \ge 0$, {\it no} barotropic model
can evolve as a pseudo-$\Lambda$ fluid; all such models evolve toward a constant nonzero density.  (It is easy to verify
that the model given in Eq. (\ref{quadratic}) with $\alpha > 1$ violates the stability constraint on the sound speed
as $\rho_{DE} \rightarrow 0$).

\subsection{Quintessence}

Now consider models in which
the dark energy is provided by a minimally-coupled
scalar field, $\phi$, with equation of motion given by
\begin{equation}
\label{motionq}
\ddot{\phi}+ 3H\dot{\phi} + \frac{dV}{d\phi} =0,
\end{equation}
where the Hubble parameter $H$ is given by
\begin{equation}
\label{H}
H = \left(\frac{\dot{a}}{a}\right) = \sqrt{\rho_T/3},
\end{equation}
and $\rho_T$ is the total density.
Since we are interested in the evolution of dark energy at relatively
late times, we will consider only the contributions of nonrelativistic
matter (baryons plus dark matter), along with the quintessence field, to $\rho_T$, and we
will ignore the contribution of radiation.

The pressure and density of the
scalar field are given by
\begin{equation}
\label{pphi}
p_\phi = \frac{\dot \phi^2}{2} - V(\phi),
\end{equation}
and
\begin{equation}
\label{rhophi}
\rho_\phi = \frac{\dot \phi^2}{2} + V(\phi),
\end{equation}
respectively, and the equation of state parameter, $w$,
is given by equation (\ref{w}).

In order to produce either asymptotic $\Lambda$ or pseudo-$\Lambda$ behavior, we need $w \rightarrow -1$, which
requires $\dot \phi \rightarrow 0$, yielding $\rho_\phi \approx V(\phi)$.  Then asymptotic $\Lambda$ evolution
requires one of two forms for the potential:  either a nonzero local minimum within which $\phi$ can settle
\cite{AS} or an asymptotically-constant value for $V(\phi)$ as $\phi \rightarrow \infty$.  The latter
potentials can arise, e.g., for potentials of the form
$V(\phi) = V_0 + V_1(\phi)$, where $V_1(\phi) \rightarrow 0$ as $\phi\rightarrow \infty$ \cite{Chang,Bag,Akrami}.

If $V(\phi)$ has no local minimum and asymptotically decays to $V(\phi) = 0$, then
pseudo-$\Lambda$ behavior can arise if $\dot \phi \rightarrow 0$
in the long-time limit.  In the nomenclature of Ref. \cite{CL}, these are ``freezing" quintessence models.  In freezing
quintessence, the value of $w$ can be initially far removed from $-1$, but $w$ approaches $-1$ as the field
rolls downhill in the potential and freezes, with $\dot \phi$ going to zero.  For a recent discussion
of freezing quintessence, see Ref. \cite{freeze}.  As we will see, $\dot \phi \rightarrow 0$ as $V(\phi) \rightarrow 0$
is a necessary but not sufficient condition for pseudo-$\Lambda$ behavior.

Freezing models were among the first types of quintessence models studied.  They arise, for example
for power-law potentials of the form \cite{RatraPeebles,Liddle,SteinhardtWangZlatev}
\begin{equation}
\label{power}
V(\phi) = V_0 \phi^{-\alpha},
\end{equation}
with $\alpha > 0$,
or for exponential potentials \cite{Wetterich,Ferreira,CLW}
\begin{equation}
V(\phi) = V_0 e^{-\lambda \phi}.
\end{equation}
In the former case, the value of $w$ is given by
\begin{equation}
1+w = \frac{\alpha}{2+\alpha},
\end{equation}
during the matter-dominated era, when the effect of the quintessence energy density on the expansion can be neglected.
At late times, when the quintessence energy begins to dominate,
$w$ decreases, asymptotically approaching $-1$,
and the universe expands as
\begin{equation}
a \sim \exp(t^{4/(4+\alpha)}).
\end{equation}
This corresponds to the intermediate inflation model \cite{inter1,inter2,inter3} discussed in the previous section.
Thus, power-law potentials of the form of Eq. (\ref{power}) yield pseudo-$\Lambda$ behavior in the limit where the scalar
field dominates the expansion.

For the exponential potential, the behavior of $w$ depends on the value of $\lambda$.  During the matter-dominated era,
the quintessence equation of state parameter tracks the matter value ($w=0$) as long as $\lambda^2 > 3$.
For $\lambda^2 < 3$, we have instead $1+w = \lambda^2/3$.  In the former case, the matter and quintessence
evolve with a constant ratio, while in the latter case, the scalar field energy density becomes the dominant
component.  However, in the latter case we not have pseudo-$\Lambda$ behavior,
because $1+w$ asymptotes to a nonzero constant, corresponding to type III
behavior.

The distinction between quintessence evolving toward asymptotic $\Lambda$ (type I) versus pseudo-$\Lambda$ (type II)
behavior is clear: it simply depends on whether $V(\phi)$ goes to zero or a nonzero constant asymptotically.
The more interesting question for quintessence is the boundary between type II and type III behavior:  when does
the scalar field give $w \rightarrow -1$ asymptotically, versus some other asymptotic value for $w$?  As
we have seen,
both negative power law potentials and the exponential potential have $\dot \phi \rightarrow 0$, but the former leads
to pseudo-$\Lambda$ evolution, while the latter
produces a value for $1+w$ that asymptotes to a nonzero constant.

To determine the conditions on the potential needed to produce pseudo-$\Lambda$ behavior, we use the equation
for the evolution
of $w$ \cite{Linder06,SS}
\begin{equation}
\label{wevol}
\frac{dw}{d \ln a} = -3 (1+w)(1-w) + \lambda(1-w)\sqrt{3 (1+w) \Omega_\phi},
\end{equation}
where we have introduced the quantity $\lambda \equiv -V^\prime/V$ and we assume
$V(\phi)$ is a decreasing function of $\phi$ with $V(\phi) \rightarrow 0$ as $\phi \rightarrow \infty$.
It is clear from Eq. (\ref{wevol}) that whenever
$\lambda \rightarrow 0$ asymptotically,
$w$ will decrease down to the limiting value of $w = -1$, giving pseudo-$\Lambda$ behavior.
This result is derived more rigorously in Refs. \cite{Macorra,Copeland2}.
Thus, the exponential potential, for which $\lambda = constant$, provides the boundary between pseudo-$\Lambda$ behavior and evolution toward
constant $w \ne -1$.

Note, however, that this result applies only in the asymptotic regime when the universe is scalar field dominated.
What happens when the universe is dominated by a separate barotropic fluid such as matter or radiation?
Can the scalar field evolve to a pseudo-$\Lambda$ state where $w \rightarrow -1$ and
$\rho_{DE} \rightarrow 0$ under these
conditions?
(Of course, in this case $\rho_{DE}$ will eventually overtake the barotropic background density, but we are
interested in the evolution before this happens).
In terms of the quintessence parameters, $w$ is given by
\begin{equation}
\label{gammaQ}
1+w = \frac{\dot \phi^2}{\dot \phi^2/2 + V(\phi)},
\end{equation}
and pseudo-$\Lambda$ behavior requires $1+w$, as well as both
the numerator and denominator in Eq. (\ref{gammaQ}) go to zero as $t \rightarrow \infty$.  But L'Hopital's theorem tells us
that if both the numerator and denominator in Eq. (\ref{gammaQ}) go to zero, then
\begin{equation}
\lim_{t \rightarrow \infty} \frac{\dot \phi^2}{\dot \phi^2/2 + V(\phi)}
= \lim_{t \rightarrow \infty}  \frac{2 \ddot \phi}{\ddot \phi + dV/d\phi},
\end{equation}
which, along with Eq. (\ref{motionq}), implies
\begin{equation}
1+w = \lim_{t \rightarrow \infty}- \frac{2 \ddot \phi}{3H \dot \phi}.
\end{equation}
Then if $1+w \rightarrow 0$, we have
\begin{equation}
\label{slowroll}
|\ddot \phi| \ll |dV/d\phi|,~ |3H\dot \phi|,
\end{equation}
which
is just the familiar slow-roll condition from the dynamics
of inflation.

For the case of a universe containing matter and quintessence, it can be shown \cite{Linder,Cahn} that most
potentials do not produce slow-roll behavior, i.e., they do not yield $|\ddot \phi| \ll |dV/d\phi|$.  (Note
that there is an ambiguity in the quintessence literature: the term ``slow-roll quintessence" is sometimes used
to refer
to scalar field evolution in a very flat potential for which $\dot \phi^2 \ll V(\phi)$, even
when Eq. (\ref{slowroll}) is not satisfied \cite{Chiba,Dutta}.  That will {\it not} be our
usage in this paper).

We can extend the results of Refs. \cite{Linder, Cahn} to show that for a universe dominated by
a barotropic fluid, there is only a single potential that yields slow-roll behavior in the sense
defined by Eq. (\ref{slowroll}), and this potential does not yield pseudo-$\Lambda$ behavior.
Consider a universe dominated by a background fluid with equation of state parameter $w_B$.
Then $H = 2/[3(1+w_B)t]$, and Eq. (\ref{motionq}) becomes
\begin{equation}
\label{motiongamma}
\ddot{\phi}+ \left(\frac{2}{(1+ w_B) t}\right)\dot{\phi} + \frac{dV}{d\phi} =0.
\end{equation}
Note that the solution to Eq. (\ref{motiongamma}) provides an expression for $\phi(t)$,
so we can write the third term in Eq. (\ref{motiongamma}) in terms of $t$ rather than $\phi$.
In particular, define the function $F(t)$ to be given by $F(t) \equiv - V^\prime(\phi(t))$, where
the prime denotes the derivative with respect to $\phi$, and our choice of sign insures that $F(t) > 0$.
Then the slow-roll condition allows us to write
\begin{equation}
\label{slowrolldotphi}
\dot \phi = \left(\frac{(1+w_B) t}{2}\right) F(t),
\end{equation}
and taking the derivative gives
\begin{equation}
\ddot \phi = \frac{1+w_B}{2} F(t) + \frac{(1+w_B) t}{2} \frac{dF(t)}{dt} \ll F(t),
\end{equation}
where the inequality is required by the slow-roll condition.
This inequality simplifies to
\begin{equation}
\label{Flimit}
\frac{1+w_B}{2}\left[1 + \frac{t}{F(t)} \frac{dF}{dt}\right] \ll 1.  
\end{equation}
Unless $w_B$ is close to $-1$ (and we will assume it is not), Eq. (\ref{Flimit}) implies
$F(t) \approx C/t$, where $C > 0$ is a constant of integration, and Eq. (\ref{slowrolldotphi})
gives $\dot \phi = C(1+w_B)/2$.  Then $\phi = C (1+w_B) t/2 + D$, with $D$ another arbitrary constant.
Combining the expressions for $\phi(t)$ and $F(t)$ yields $dV/d\phi =  - C^2 (1+w_B)/2 (\phi-D)$, so
\begin{equation}
V(\phi) = V_0 - \frac{C^2 (1+w_B)}{2} \ln(\phi-D),
\end{equation}
with $V_0$ another arbitrary constant.
Thus, the only potential that produces slow-roll behavior when the expansion is background dominated is
the logarithmic potential.  However, this manifestly does not produce
pseudo-$\Lambda$ behavior:  this solution gives a constant value for $\dot
\phi$, while $V(\phi)$ is a decreasing function of $\phi$, so Eq. (\ref{gammaQ})
indicates that $w$ increases with time, rather than decreasing
asymptotically to zero.

\section{Relation to $w < -1$ models}

If $w < -1$, then the weak energy condition is violated, and the dark energy density
increases as the universe expands. This possibility
was first proposed by Caldwell \cite{Caldwell}, who dubbed it phantom
dark energy, and it has been extensively explored since then.
Constant-$w$ models for which $w < -1$ lead generically to a big rip, in which
the density and scale factor both become infinite at a finite time $t_r$.

If $w < -1$, but $w \rightarrow -1$ asymptotically, the situation is more complicated.
In this case, there are three different possibilities.  The first is a standard big rip, with
$\rho_{DE} \rightarrow \infty$
as $t \rightarrow t_r$.  However, one can also have a little rip, for which
$\rho_{DE} \rightarrow \infty$ as $t \rightarrow \infty$ \cite{littlerip,littlerip2},
or a pseudo-rip, which has $\rho_{DE} \rightarrow constant$ as $t \rightarrow \infty$
\cite{pseudorip}.

Here we note the duality between these phantom models and the asymptotic $\Lambda$ and pseudo-$\Lambda$
models.
Consider the evolution of $\rho_{DE}$ for $w > -1$ given
by Eq. (\ref{integral}).  If we replace $1+w$ with $-(1+w)$, then $\rho_{DE}$ maps
to $1/\rho_{DE}$, and the models corresponding to asymptotic $\Lambda$ behavior and pseudo-$\Lambda$
behavior map to the pseudo-rip and the little/big rip, respectively.  Thus,
the boundary between asymptotic $\Lambda$ and pseudo-$\Lambda$ behavior for $1+w > 0$ corresponds
to the boundary between the pseudo-rip and the little or big rip for $1+w < 0$.  This can be seen
explicitly in Ref. \cite{littlerip2}, which derives the condition on $w(a)$ that distinguishes between
the pseudo-rip and the little or big rip.  This condition is exactly the same as our condition on the integral
in Eq. (\ref{integral}); when $1+w < 0$ and this integral converges, $\rho$ is asymptotically constant,
and we have a pseudo-rip, while when it diverges, $\rho_{DE} \rightarrow \infty$, and we have a little
or big rip.

\section{Discussion}

Dark energy with $w \rightarrow -1$ does not correspond to a single evolutionary behavior for $\rho_{DE}$; instead,
it can describe models for which $\rho_{DE} \rightarrow constant$ (asymptotic $\Lambda$) or $\rho_{DE} \rightarrow 0$
(pseudo-$\Lambda$).  Clearly, it is possible to produce models of both types that are arbitrarily similar
to each other (and to $\Lambda$CDM) at the present, while yielding wildly different predictions for the future
evolution of the universe:  asymptotic $\Lambda$ models always evolve toward exponential expansion, while
pseudo-$\Lambda$ models produce subexponential future expansion.
This is not surprising, as a similar result was noted in Ref. \cite{Ash} for
models with $w < -1$; models can be arbitrarily similar to $\Lambda$CDM, while diverging in the future
into big-rip, little-rip, or pseudo-rip final states.

Given the difficulty of distinguishing observationally between the two
classes of models that we have examined in this paper, our results are probably more
important for what they tell us about the limitations of using $w$ alone to parametrize dark energy.
The observable quantity that distinguishes dark energy models is $\rho(a)$, or equivalently, $H(a)$.
In terms of $\rho(a)$, the asymptotic $\Lambda$ and pseudo-$\Lambda$ models are completely
different types of models.  However, they both map onto the same asymptotic value of $w$, which happens
to be the value favored by current observational data.

\end{document}